\documentclass[useAMS,usenatbib,usegraphicx]{mn2e}
\usepackage{psfig}   
\usepackage{graphicx}
\usepackage{amssymb}
\usepackage{subfigure}

\title[Mass segregation is not energy equipartition]{Mass segregation in star clusters is not energy equipartition}

\author[R.~J.~Parker et al.]{
  Richard J.~Parker$^{1}$\thanks{E-mail: R.J.Parker@ljmu.ac.uk}, Simon
  P.~Goodwin$^2$, Nicholas J.~Wright$^{3}$, Michael R. Meyer$^4$ \newauthor \hspace*{0.03cm} and Sascha P. Quanz$^4$
  \vspace*{0.1cm}\\
   $^1$ Astrophysics Research Institute, Liverpool John Moores University, 146 Brownlow Hill, Liverpool, L3 5RF, UK \\
  $^2$ Department of Physics and Astronomy, University of Sheffield,
    Sheffield, S3 7RH, UK\\
   $^3$ Astrophysics Group, Keele University, Keele, Staffordshire,
   ST5 5BG, UK\\
  $^4$ Institute for Astronomy, ETH Z{\"u}rich, Wolfgang-Pauli-Strasse 27, 8093, Z{\"u}rich, Switzerland}

\begin{document}

\date{}
                             
\pagerange{\pageref{firstpage}--\pageref{lastpage}} \pubyear{2016}

\maketitle

\label{firstpage}

\begin{abstract}
Mass segregation in star clusters is often thought to indicate the
onset of energy equipartition, where the most massive stars impart
kinetic energy to the lower-mass stars and brown dwarfs/free floating
planets. The predicted net result of this is that the centrally
concentrated massive stars should have significantly lower velocities
than fast-moving low-mass objects on the periphery of the cluster. We
search for energy equipartition in initially spatially and kinematically
substructured $N$-body simulations of star clusters with $N = 1500$
stars, evolved for 100\,Myr. In clusters that show significant mass segregation we find no differences in the proper motions or radial velocities as a function of mass. The kinetic energies of all stars decrease as the clusters relax, but the kinetic energies of the most massive stars do not decrease faster than those of lower-mass stars. These results suggest that dynamical mass segregation -- which is observed in many star clusters -- is not a signature of energy equipartition from two-body relaxation.    
\end{abstract}


\section{Introduction}

The majority of star formation occurs in regions that exceed the mean
density of the Galactic disc by several orders of magnitude
\citep{Blaauw64,Lada03,Porras03,Bressert10}. A fraction of these
star-forming regions subsequently form bound, centrally concentrated
star clusters \citep{Kruijssen12b,Parker14b}, whose occurrence rate
depends on their Galactic environment \citep{Adamo15}. Understanding
the subsequent dynamical evolution of clusters can place constraints
on the initial conditions of star formation, and the likely birth
environment of the majority of stars in the Galaxy.

One observed characteristic of star clusters is the relative spatial distribution of the most massive stars compared to low-mass stars. The over-concentration of massive stars in the cluster centre, referred to as `mass segregation', is either a primordial outcome of the star formation process \citep[e.g.][]{Zinnecker82,Bonnell97} or a later dynamical effect \citep{Allison09b}. 

In either scenario, mass segregation is often assumed to be the first
signature of energy equipartition in clusters, in which all stars
have the same kinetic energy. In this picture, the most massive stars
exchange kinetic energy with low-mass stars as they move to the centre
of the cluster and slow down, and the low-mass stars (and/or brown
dwarfs and free floating planets) gain kinetic energy and are ejected
to the outskirts. 

When full energy equipartition occurs the velocity dispersion,
  $\sigma$ of every subset of stars is
  proportional to the average stellar mass $m$ in the subset,  
\begin{equation}
\sigma \propto m^{-0.5}.
\end{equation}
Before this occurs, a cluster is expected to attain partial energy equipartition where the more massive stars have lower velocities than average-mass stars.

The timescale for energy equipartition is many relaxation times,
$t_{\rm relax}$ \citep{Spitzer69}  where
\begin{equation}
t_{\rm relax} = \frac{N}{8 {\rm ln} N}t_{\rm cross},
\end{equation}
and $N$ is the number of stars and $t_{\rm cross}$ is the crossing
time. For a typical crossing time in a dense cluster of 0.1\,Myr,
$t_{\rm relax} \sim 100$\,Myr for $N = 1000$ stars and as a consequence would only be
expected in the oldest clusters (globular and old open
clusters). Several studies \citep{Spitzer69,Vishniac78,Khalisi07} have shown that full energy equipartition
never occurs in globular clusters, but these systems sometimes reach partial energy equipartition \citep{Trenti13,Sollima15,Bianchini16}. 

It is currently unclear whether younger, less massive clusters exhibit
energy equipartition. Their lower masses (and hence number of stars) relative to globular clusters
means that their relaxation times will be much shorter, and many of
them display prominent mass segregation of the most massive stars
\citep[e.g.][]{Hillenbrand98,Gouliermis09}. Furthermore,
\citet{Allison09b,Allison10,Parker14b} show that more realistic spatially and
kinematically substructured initial conditions accelerate mass
segregation in clusters, but it is unclear if these initial conditions
also lead to (partial) energy
equipartition.

In this Letter, we follow the dynamical evolution of clusters to an age of 100\,Myr using $N$-body simulations to investigate whether any mass segregation that occurs can be attributed to energy equipartition. We describe the simulations in Section~\ref{method}, we present our results in Section~\ref{results}, we provide a discussion in Seection~\ref{discuss} and we conclude in Section~\ref{conclude}.

\section{Method}
\label{method}

We follow the formation and evolution of our model clusters using
$N$-body simulations. Observations of young star-forming regions, and
the giant molecular clouds from which they form have a hierarchical
and substructured morphology irrespective of their mass \citep{Cartwright04,Sanchez09,Walker15}. Furthermore, the velocity dispersions of
star-forming cores within filaments are sub-virial \citep{Peretto06,Kauffmann13}, and kinematic
substructure is observed in young star-forming regions \citep{Hacar13,Alfaro16}. 

We therefore set up our $N$-body clusters with both spatial and
kinematic substructure, using the fractal generator from
\citet{Goodwin04a}. This determines the amount of spatial and
kinematic substructure from one parameter, the fractal dimension
$D$. We then scale the velocities of the stars so that the whole
region is subvirial with a virial ratio $\alpha_{\rm vir} = 0.3$ (virial equilibrium is $\alpha_{\rm vir} = 0.5$) -- i.e.\,\,it will collapse to form a cluster
\citep{Allison10,Parker14b,Parker16b}. 

The fractal clusters have 1500 stars (similar to the lowest-mass open
clusters -- i.e.\,\,those with the shortest relaxation times) with masses drawn from a \citet{Maschberger13} IMF between 0.01\,M$_\odot$ and 50\,M$_\odot$ and
an initial radius of 1\,pc. The fractal dimension is $D = 1.6$, which
gives a very substructured initial distribution \citep[and leads to the most
pronounced dynamical mass segregation,][]{Allison10}. We also ran a set of
simulations containing primordial binaries with properties similar to
systems in the Galactic field \citep{Raghavan10,Reggiani11a}. 


We ran 20 versions of the initial conditions, identical apart from the
random number seed used to initialise the stellar masses, positions
and velocities. The $N$-body simulations were evolved for 100\,Myr
(the typical relaxation time of these clusters) using the \texttt{kira}
integrator within the Starlab environment \citep{Zwart99,Zwart01}. We
implement stellar evolution using the \texttt{SeBa} look-up tables \citep{Zwart96},
which are also part of Starlab. 

\section{Results}
\label{results}

\subsection{Cluster definition}

The fractal simulations initially erase their substructure and collapse to
form a smooth, centrally concentrated cluster. The presence of
substructure, in tandem with correlated velocities on local scales,
has been shown to facilitiate dynamical mass segregation at very early
times ($\sim$ 1\,Myr) \citep{Allison09b,Allison10}. During this
violent relaxation process, unstable multiple systems consisting of
the most massive stars form in the centre of clusters
\citep{Allison11}. However, these Trapezium-like systems are unstable
and can lead to the ejection of one or more massive stars.   

Furthermore,  as the clusters evolve over the 100\,Myr, the process of two-body
relaxation leads to further ejections of both massive \citep{Oh15} and
lower-mass stars. When stars are ejected at high velocities
($>$10\,km\,s$^{-1}$) they can travel several tens of pc during the
simulation and are unlikely to be observationally associated with the
star cluster. 

For this reason, we consider the star cluster boundary to be twice 
the half-mass radius, $r_{1/2}$ of all stars; this encompasses most of the stars
in the cluster but disregards the ejected stars that have travelled
beyond the periphery of the cluster. Defining the cluster boundary
  using the position of the furthest energetically bound star from the
cluster centre gives very similar results to using $2r_{1/2}$ \citep{Parker12a}. In the example simulation we present here, of the 1500 stars in the cluster initially, 1039 remain at the
  end of this simulation. 

\subsection{Mass segregation} 

We define mass segregation in two different ways. First, we use the
$\Lambda_{\rm MSR}$ mass segregation ratio from \citet{Allison09a},
which compares the relative spatial distributions of a chosen subset
of stars (e.g.\,\,the 10 most massive) to randomly chosen subsets. A
minimum spanning tree (MST) is used to quantify the typical length between
the most massive stars $l_{\rm sub}$, and this is compared to the mean MST length of
many realisations of randomly chosen stars $\langle l_{\rm average} \rangle$ (which may or may not
include members of the most massive subset):
\begin{equation}
\Lambda_{\rm MSR} = {\frac{\langle l_{\rm average} \rangle}{l_{\rm
      sub}}} ^{+ {\sigma_{\rm 5/6}}/{l_{\rm sub}}}_{- {\sigma_{\rm
      1/6}}/{l_{\rm sub}}}.
\label{lambda_eq}
\end{equation}
$\Lambda_{\rm MSR} = 1$ indicates no mass segregation, whereas
$\Lambda_{\rm MSR} >> 1$ indicates strong segregation. The lower
(upper) uncertainty is defined as the MST
length which lies 1/6 (5/6) of the way through an ordered list of all
the random lengths (corresponding to a 66 per cent deviation from  the
median value, $\langle l_{\rm average} \rangle$).

\begin{figure*}
  \begin{center}
\setlength{\subfigcapskip}{10pt}
\hspace*{-1.5cm}\subfigure[Evolution of $\Lambda_{\rm MSR}$]{\label{lambda-a}\rotatebox{270}{\includegraphics[scale=0.35]{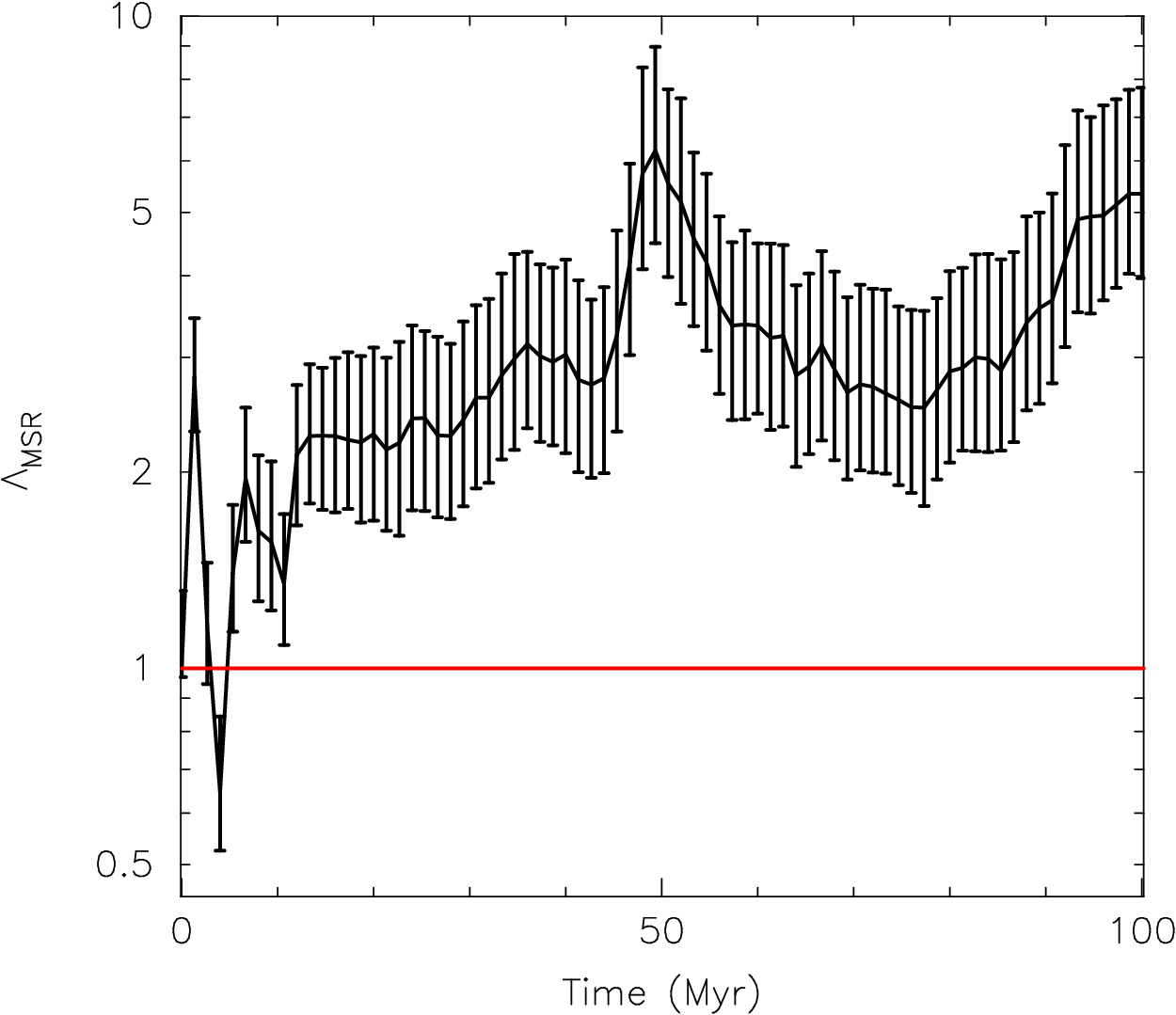}}}
\hspace*{0.3cm} 
\subfigure[$\Lambda_{\rm MSR}$; 100\,Myr]{\label{lambda-b}\rotatebox{270}{\includegraphics[scale=0.35]{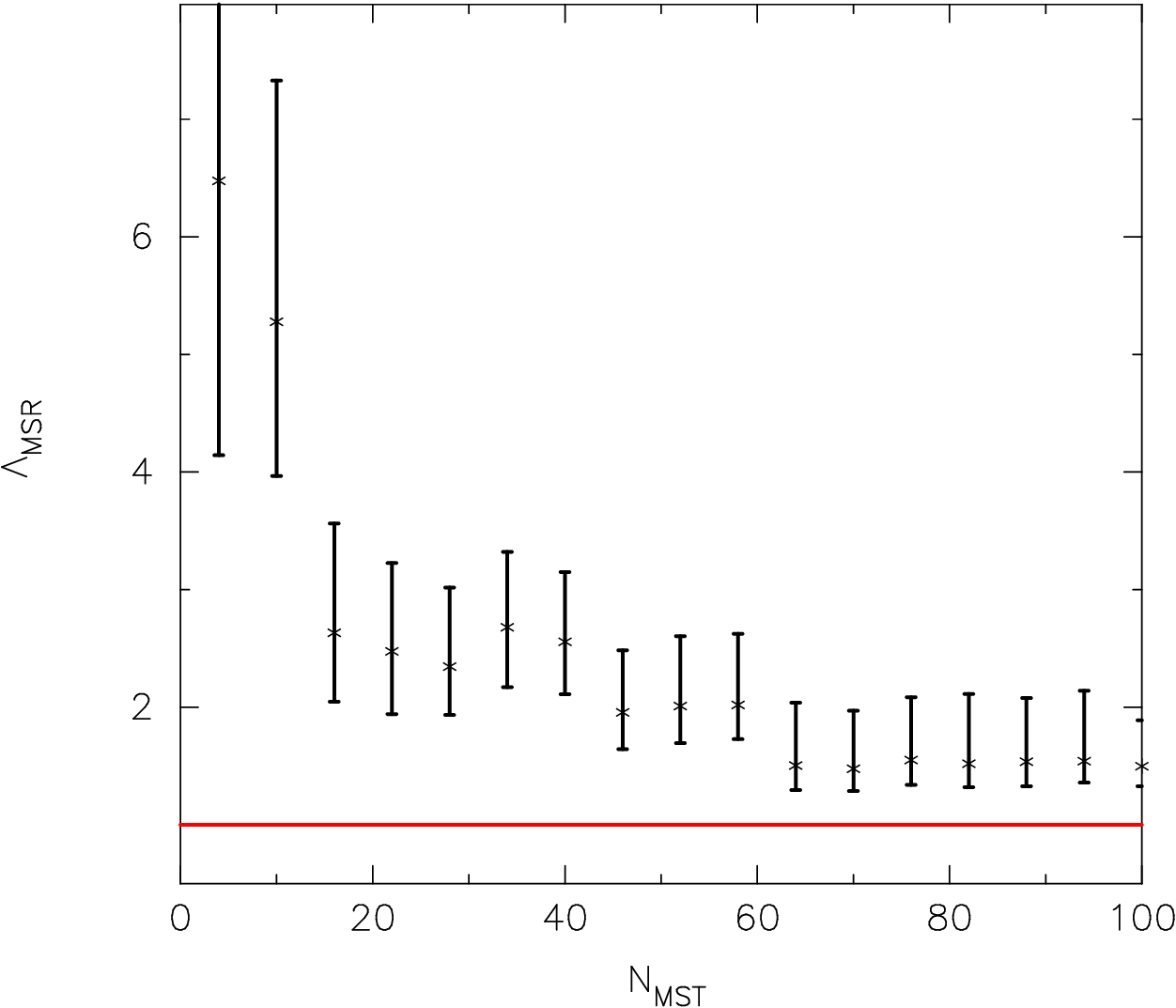}}}
\caption[bf]{Mass segregation in the simulations as defined by the
  $\Lambda_{\rm MSR}$ ratio, with uncertainties defined by Eqn.~\ref{lambda_eq}. In panel (a) we show the  evolution of $\Lambda_{\rm MSR}$ for the 10 most massive
  stars in the simulation. The cluster rapidly dynamically mass
  segregates, before ejecting some of the most massive stars at around
5\,Myr. Once these stars have travelled beyond twice the half-mass
radius they are excluded from the determination of $\Lambda_{\rm
  MSR}$, and the cluster remains mass segregated until 100\,Myr. In
panel (b) we show 
$\Lambda_{\rm MSR}$ as a function of the $N_{\rm MST}$ most massive
stars at 100\,Myr. The cluster is mass segregated down to the
$\sim$50$^{\rm th}$ most massive star. }
\label{lambda}
  \end{center}
\end{figure*}

In Fig.~\ref{lambda-a} we show the evolution of $\Lambda_{\rm MSR}$
for the ten most massive stars (3.4 -- 5.2\,M$_\odot$) in a cluster that shows behaviour
typical of the full suite of simulations. As in
\citet{Allison10,Parker14b} dynamical mass segregation occurs early in
the simulation, but ejections of the most massive stars cause the
signal to decay, before the ejected massive stars move beyond the
cluster limits and are not included in the determination. A strong mass
segregation signal returns, which is maintained even as the most
massive stars lose mass due to stellar evolution\footnote{We also ran
  a control simulation with no stellar evolution and found very similar
  results.}. 

 We show the level of mass segregation at 100\,Myr in
Fig.~\ref{lambda-b}. The plot shows $\Lambda_{\rm MSR}$ as a function
of the $N_{\rm MST}$ most massive stars, and the cluster is clearly
mass segregated down to the 50$^{\rm th}$ most massive star which has a mass 1.05\,M$_\odot$, although
stochastic differences in evolution mean that other clusters can be
mass segregated to fewer, or more stars.

\begin{figure}
\begin{center}
\rotatebox{270}{\includegraphics[scale=0.37]{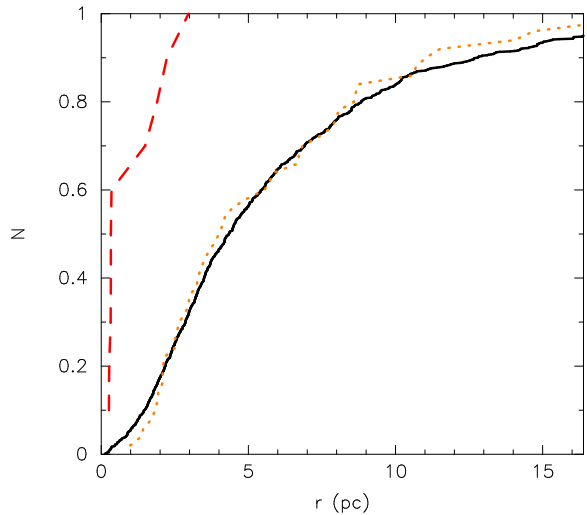}}
\end{center}
\caption[bf]{The radial distribution of the ten most
massive stars (shown by the red dashed line), all stars (the solid
black line) and the brown dwarfs (dotted orange line) at 100\,Myr. The
most massive stars are clearly more centrally concentrated, but the
brown dwarfs are not more distributed than the average star.}
\label{radial_mf}
\end{figure}

In a cluster that no longer has primordial substructure and is mass segregated, we would
expect a clear difference in the cumulative distributions of the
positions of the most massive stars compared to the cluster
average. If the cluster has also undergone energy equipartition, we
might expect that the lowest mass objects (free floating planets and
brown dwarfs) to be further out than the stars. 

In Fig.~\ref{radial_mf} we show the cumulative radial distribution of the
massive stars by the red dashed line, all objects by the solid black
line, and the brown dwarfs ($m < 0.08$M$_\odot$) by the dotted orange line. Whilst the most
massive stars are more centrally concentrated than the cluster
average, the brown dwarfs are not further from the cluster centre than
the average objects.

\begin{figure}
\begin{center}
\rotatebox{270}{\includegraphics[scale=0.37]{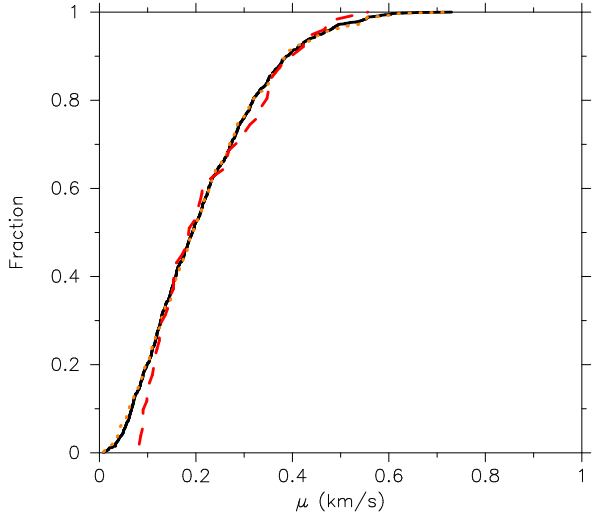}}
\end{center}
\caption[bf]{The proper
motions of the ten most massive stars (red dashed line), all stars
(solid black line) and the brown dwarfs (orange dotted line) at
100\,Myr. All objects have the same proper motion velocity distribution.}
\label{proper_motions}
\end{figure}

\subsection{Proper motions}

\begin{figure}
  \begin{center}
\setlength{\subfigcapskip}{10pt}
\rotatebox{270}{\includegraphics[scale=0.37]{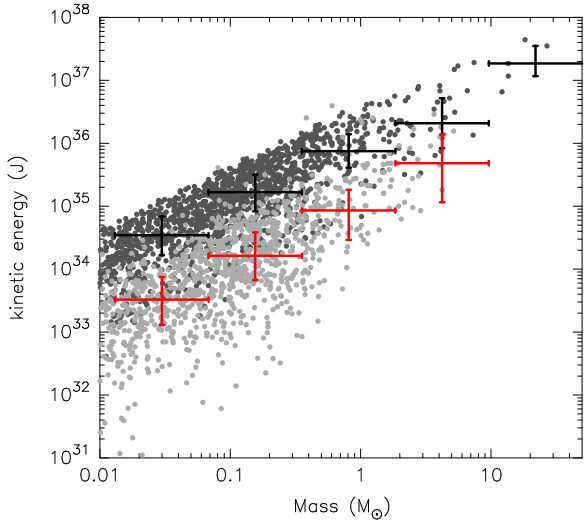}}
\caption[bf]{Kinetic energy as a function of stellar mass for the
  initial conditions (dark grey points) and at 100\,Myr
  (light grey points). The median
  kinetic energies for equally-spaced mass intervals are shown by the
  error bars, where the horizontal error bars indicate the mass range, and
  the vertical error bars are the interquartile range of
kinetic energies. The black error bars are for the initial
  conditions; the red error bars are the values at 100\,Myr.}
\label{ke_mbins}
  \end{center}
\end{figure}

\begin{figure}
\begin{center}
\rotatebox{270}{\includegraphics[scale=0.37]{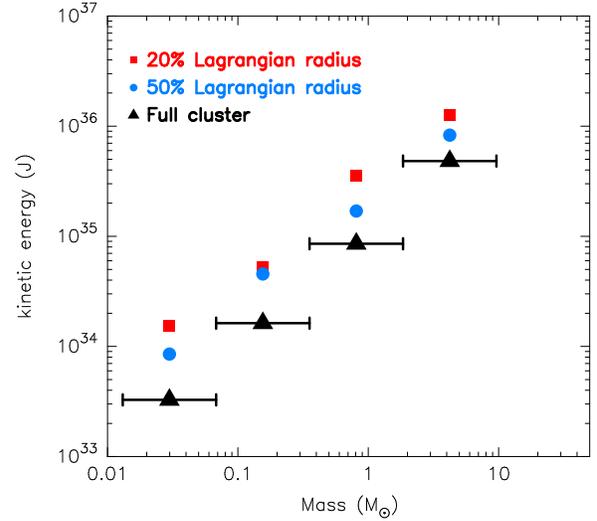}}
\end{center}
\caption[bf]{The median kinetic energies for equally spaced mass
  intervals (indicated by the horizontal error bars) at 100\,Myr for different Lagrangian radii. }
\label{ke_mbins_lagrange}
\end{figure}

If energy equipartition has occurred in a star cluster, we would
expect the most massive objects to be moving more slowly than the
average cluster stars, and that the lowest mass objects (brown dwarfs
and free-floating planets) would be moving faster. In
Fig.~\ref{proper_motions} we show the cumulative distribution of the
proper motions of stars in the cluster. We mimic an observational
determination by comparing the change in positions in the \emph{x-y}
plane between timesteps ($\Delta t = 0.1$\,Myr) in the simulation. We also checked that we obtain the same result using the respective $v_x$ and $v_y$ velocities.

Fig.~\ref{proper_motions} clearly shows that the most massive stars are not moving more slowly, nor are the brown dwarfs moving faster than the average stars in the cluster. We note that \citet{Osorio14} find evidence that the proper motions of objects in the Plaiedes cluster increase with decreasing mass. We suggest that this result is probably due to the stochastic nature of cluster dynamics; \citet{Parker14c} find that dynamical evolution can cause brown dwarfs to be less spatially concentrated than higher-mass objects, but this is a stochastic effect that only occurs in 20\% of their simulations.

\subsection{Kinetic energies}

In order to directly look for energy equipartition, we plot the
kinetic energies of individual stars as a function of their mass in
Fig.~\ref{ke_mbins}. We show the individual kinetic energies for the
stars before dynamical evolution ($t = 0$\,Myr, the dark grey
  points) and at the end of the simulation ($t = 100$\,Myr, the
  light grey points). We also show the median kinetic energy in a
series of mass bins, where the vertical error bar is the interquartile
range of kinetic energies. Black error bars are for the initial
  values, red are the values at 100\,Myr. Due to stellar evolution, the rightmost bin is not present after 100\,Myr.

As the cluster relaxes and expands, each star slows down and hence
loses kinetic energy on average, but the rate at which energy is lost
is independent of stellar mass. There is no indication of energy
equipartition, and the stars that are mass segregated are not distinct
in this plot. In Fig.~\ref{ke_mbins_lagrange} we show the median binned kinetic
  energies at 100\,Myr for different Lagrangian radii. There is no
  radial dependence on the kinetic energy distribution. We also
determined the velocity dispersion as a function of mass for the mass
bins in this plot (as is more commonly done for globular cluster
simulations). We find that at all ages the velocity dispersions remain
constant (within a few per cent) as a function of mass, and at
  different Lagrangian radii. 


\section{Discussion}
\label{discuss}

Our $N$-body simulations show that where mass segregation is
pronounced in low-mass, intermediate age star clusters, there is no
indication of the onset of energy equipartition. If mass segregation
was indicative of energy equipartition, we would expect the most
massive stars to be moving more slowly, as their kinetic energies
would be the same as those of lower-mass stars. We find that the most massive
stars do not slow down faster than the average stars during mass segregation. In some ways this is not surprising; as
the massive stars move towards the cluster centre, they fall deeper
into the gravitational potential and would be expected to speed
up. Once in the cluster centre, they often behave as a separate
sub-system \citep{Allison11}, which relaxes by ejecting one or more of
the massive stars from the centre.

This is a somewhat continuous or self-regulating process. The level of
mass segregation is generally constant throughout the simulation
(Fig.~\ref{lambda-a}); if a star is ejected or loses mass through
stellar evolution, it is replaced in the potential well by the next
most massive star such that the highest mass member in the subset of the ten
most massive stars that are significantly mass segregated decreases
from 30\,M$_\odot$ to 7\,M$_\odot$ during the lifetime of the
simulation.

\citet{Spitzer69} and \citet{Vishniac78} show analytically that the formation of a
sub-system of the most massive stars in the core leads to the
suppression of energy equipartition in clusters. Our initial
conditions, which contain spatial and kinematic substructure, are
informed by observations of young star-forming regions and lead to
mass segregation and the formation of massive star sub-systems
\citep{Allison11} on faster timescales than for more commonly adopted \citet{Plummer11} or
\citet{King66} profiles. If open clusters formed from initial
conditions similar to observed star-forming regions, we therefore would not
expect any energy equipartition to occur through dynamical evolution.   

We repeated the simulations without stellar evolution and find similar
results; the most massive stars sink to the centre more rapidly than
the average star and form an unstable higher-order multiple system,
which decays by ejecting one or more massive stars. This suggests that
mass-loss via stellar evolution is not a strong influence on our
results.

In the simulations that included primordial
binaries, their presence suppresses the level of mass segregation measured by
$\Lambda_{\rm MSR}$, and it occurs only half
as often as in clusters containing all single stars. In the clusters
where it does occur, we find no evidence for energy equipartition. 

We do not find any clear evidence in the simulations that the massive
stars are imparting energy to lower-mass objects. Several observations
have shown that lower-mass objects (brown dwarfs and free-floating
planets) appear to be more sparsely concentrated, and are moving at
faster velocities, than the average stars in a cluster
\citep{Andersen11,Kumar07,Caballero08,Osorio14}. We suggest that this
could be a stochastic effect from dynamical evolution of a dense
cluster which occasionally preferentially ejects the lowest mass
objects to the outskirts of the cluster \citep{Parker14c}. However, this happens in only 20\,per cent of
simulated clusters, and free-floating planets have also been shown to tend to
move with similar velocities to the stellar members of the cluster
\citep{Parker12a}.

\section{Conclusions}
\label{conclude}

We perform $N$-body simulations of the evolution of $N = 1500$ clusters with stellar evolution for the first 100\,Myr of their lifetimes. We look for mass segregation (an over concentration of the most massive stars with respect to the average stars) and when it occurs, look for evidence for energy equipartition. Our conclusions are as follows. 

(i) Most clusters reach mass segregation on timescales of 10\,Myr (100 crossing times), and maintain the level of mass segregation for the duration of the simulation. 

(ii) The clusters are significantly mass segregated  down to the
$\sim$50$^{\rm th}$ most massive star (out of 1039 stars which remain in the cluster). Stellar evolution and dynamical
ejections can change the identities of the stars that are mass
segregated, but only occasionally cause the mass segregation signature
to disappear. The presence of primordial binaries suppresses mass
segregation and we will explore this in a future paper.

(iii) The stars that are mass segregated are not moving with slower velocities than the average stars, which would be expected if they were attaining (partial) energy equipartition. 

(iv) When we look at the individual kinetic energies of stars as a
function of stellar mass, we see no evidence that the most mass
segregated stars have kinetic energies that are equal to, or even
  tend towards those of average-mass, or
low-mass stars, and the kinetic energy decreases for all stars as the
clusters relax. 

In summary, we suggest that any mass segregation observed in young
($<$10\,Myr) and intermediate age (10 -- 500\,Myr) clusters is not a signature of energy equipartition. Rather, it is simply either primordial mass segregation from the outcome of star formation, or dynamical mass segregation due to violent, and/or two-body relaxation. 

\section*{Acknowledgements}

We thank the anonymous referee for a prompt and helpful report. RJP acknowledges support from the Royal Astronomical Society in the form of a research fellowship.

\bibliography{general_ref}

\label{lastpage}

\end{document}